\documentclass[reprint,amssymb,amsmath,superscriptaddress,aps]{revtex4-2}
\pdfoutput=1

\usepackage{graphicx,soul}
\usepackage{bm}
\usepackage{bbm}
\usepackage{amsmath}
\usepackage{amsthm}
\usepackage{amssymb}
\usepackage{tabularx}
\usepackage{gensymb}
\usepackage{amssymb}
\usepackage{amsmath}
\usepackage{graphicx}
\usepackage{dcolumn}
\usepackage[usenames, dvipsnames]{color}
\usepackage[caption=false]{subfig}
\usepackage{bm}
\usepackage[colorlinks=true,linkcolor=blue,citecolor=blue,urlcolor=blue,breaklinks=true]{hyperref}

\definecolor{ngreen}{rgb}{0.2,0.6,0.2}
\definecolor{amethyst}{rgb}{0.6, 0.4, 0.8}
\definecolor{jdarkblue}{rgb}{0.1, 0.45, 0.75}

\definecolor{golden}{rgb}{0.75,0.6,0.15}

\DeclareMathAlphabet\mathbfcal{OMS}{cmsy}{b}{n}

\newcommand{\ket}[1]{|#1\rangle}

\begin{document}



\title{Entanglement distillation rates exceeding the direct transmission bound}

\author{Farzad \surname{Ghafari}}
\email{f.ghafari@griffith.edu.au}
\affiliation{Centre for Quantum Dynamics, Griffith University, Yuggera Country, Brisbane, Queensland 4111, Australia}
\affiliation{Centre for Quantum Computation and Communication Technology (Australian Research Council), Australia}
\author{Josephine Dias}
\affiliation{Centre for Quantum Computation and Communication Technology (Australian Research Council), Australia}
\affiliation{School of Mathematics and Physics, University of Queensland, Brisbane 4072, Australia}
\author{L Krister Shalm}%
\author{Varun B Verma}%
\affiliation{National Institute of Standards and Technology, 325 Broadway, Boulder, Colorado 80305, USA.}
\author{Sergei Slussarenko}%
\email{s.slussarenko@griffith.edu.au}
\affiliation{Centre for Quantum Dynamics, Griffith University, Yuggera Country, Brisbane, Queensland 4111, Australia}
\affiliation{Centre for Quantum Computation and Communication Technology (Australian Research Council), Australia}
\author{Timothy C Ralph}%
\affiliation{Centre for Quantum Computation and Communication Technology (Australian Research Council), Australia}
\affiliation{School of Mathematics and Physics, University of Queensland, Brisbane 4072, Australia}
\author{Geoff J Pryde}%
\affiliation{Centre for Quantum Dynamics, Griffith University, Yuggera Country, Brisbane, Queensland 4111, Australia}
\affiliation{Centre for Quantum Computation and Communication Technology (Australian Research Council), Australia}

\date{\today}

\maketitle


{\bf Entanglement distribution is crucial for quantum communication and cryptography~\cite{briegel1998,liao2018,zukowski1993,wehner2018} but is hindered by channel loss and decoherence. Noiseless linear amplification (NLA)~\cite{ralph2008nondeterministic, Xiang2010} is a probabilistic protocol that supports noiseless amplification without violating the no-cloning theorem, aiding in tasks like entanglement distillation and enhanced metrology. The probabilistic nature of NLA and other quantum repeater elements creates a significant resource overhead, which depends on the success rate of each individual probabilistic step~\cite{briegel1998}. We experimentally demonstrate a technique to increase NLA success probability while maintaining the amplification gain and the fidelity of the amplified state with a maximally entangled state. We show that symmetrically distributing loss before and after the amplification provides an improved scaling of the success rates of NLA. Our results are a critical step towards scalable quantum repeaters and enable efficient long-distance quantum communication.}

Entanglement degrades when distributed between remote parties due to loss and decoherence in quantum channels, making direct data transmission impractical for real-world applications~\cite{Vidal2002,Gisin2007,Eisert2002}. For instance, typical loss in standard telecommunication optical fibers, such as single-mode fibers (SMF), is around 0.2 to 0.25 dB/km at a wavelength of 1550 nm~\cite{Agrawal2002}. Unlike classical communication, where noiseless signal amplification can be performed deterministically, quantum communication cannot rely on classical amplification techniques. In quantum mechanics, the no-cloning theorem prohibits the deterministic, noise-free amplification of an unknown quantum state~\cite{Wootters1982}. Therefore, probabilistic methods like entanglement distillation~\cite{Xiang2010,dong2008experimental}--—which allows two parties to extract a smaller number of highly entangled states from a larger set of less entangled ones using local operations and classical communication~\cite{Bennett1996}—--and the development of quantum repeater technology~\cite{briegel1998} are essential for achieving robust quantum communication.

An example of a heralded probabilistic protocol that can distil entanglement is NLA~\cite{ralph2008nondeterministic, Xiang2010}. Although inherently probabilistic, NLA schemes enable successful amplification events to be heralded by an independent signal, such as the detection of an ancilla photon~\cite{Zavatta2011,Ferreyrol2010}. When applied to one arm of an entangled state they can herald a probabilistic increase in entanglement, making them suitable for integration into larger frameworks like quantum repeaters. NLA has demonstrated its crucial role in quantum optics, with numerous experimental implementations~\cite{Xiang2010,Ferreyrol2010}, including applications such as entanglement distillation~\cite{Feng2019}, entanglement purification~\cite{Sangouard2008,Gisin2010}, qubit amplification~\cite{Kocsis2013}, channel amplification~\cite{ulanov2015undoing,Slussarenko2021}, and enhanced metrology~\cite{Caves2012}. 

An essential consideration regarding entanglement distillation is its probability of success. If we consider dual-rail photonic qubits being transferred over a lossy channel then it can be shown that any entanglement distillation protocol that purifies the state will work with a probability of success, $p$, that (at best) is equal to the probability that the photon would have been transmitted through the bare channel \cite{pirandola2017fundamental}. This limit means that in any quantum repeater scenario where each node performs an entanglement distillation process, there will be no increase in overall efficiency with which entanglement is transmitted unless quantum memories are used to hold entanglement in the nodes. Further, if quantum memories are employed to hold successfully distilled copies of entanglement, then $p$ sets the requirements for the memory's minimum hold time. Therefore, any effort to increase $p$ is of great importance to quantum communication.

Surprisingly, examples of single-rail photonic quantum communications protocols which break this limit without requiring quantum memory have been found in recent research. These include twin-field quantum key distribution (QKD) \cite{lucamarini2018overcoming}, continuous variable QKD \cite{winnel2021overcoming} and single-rail entanglement distribution \cite{caspar2020heralded,Mauron2022}. The latter two examples propose the use a “distributed” version of NLA.

Here, we experimentally demonstrate an advanced technique to increase the probability of success of entanglement distillation using NLA without compromising amplification gain or the fidelity of the amplified state. Building on the theoretical framework proposed by Mauron and Ralph~\cite{Mauron2022}, we show that symmetrically distributing the loss before and after the amplification stage significantly improves the probability of success compared to conventional NLA protocols, where the amplifier is positioned at the end of a lossy channel. We specifically demonstrate that in a high-loss regime, distributed NLA distils entanglement and overcomes the probability of success bound of direct transmission, while conventional NLA falls significantly below the bound. This development is a critical step towards scalable quantum repeaters and other quantum communication applications, enabling more efficient and reliable quantum information transfer over long distances.

\section{Materials and methods}

\subsection{NLA protocol} 
  \begin{figure*}
	\centering
	\includegraphics[width=1\linewidth]{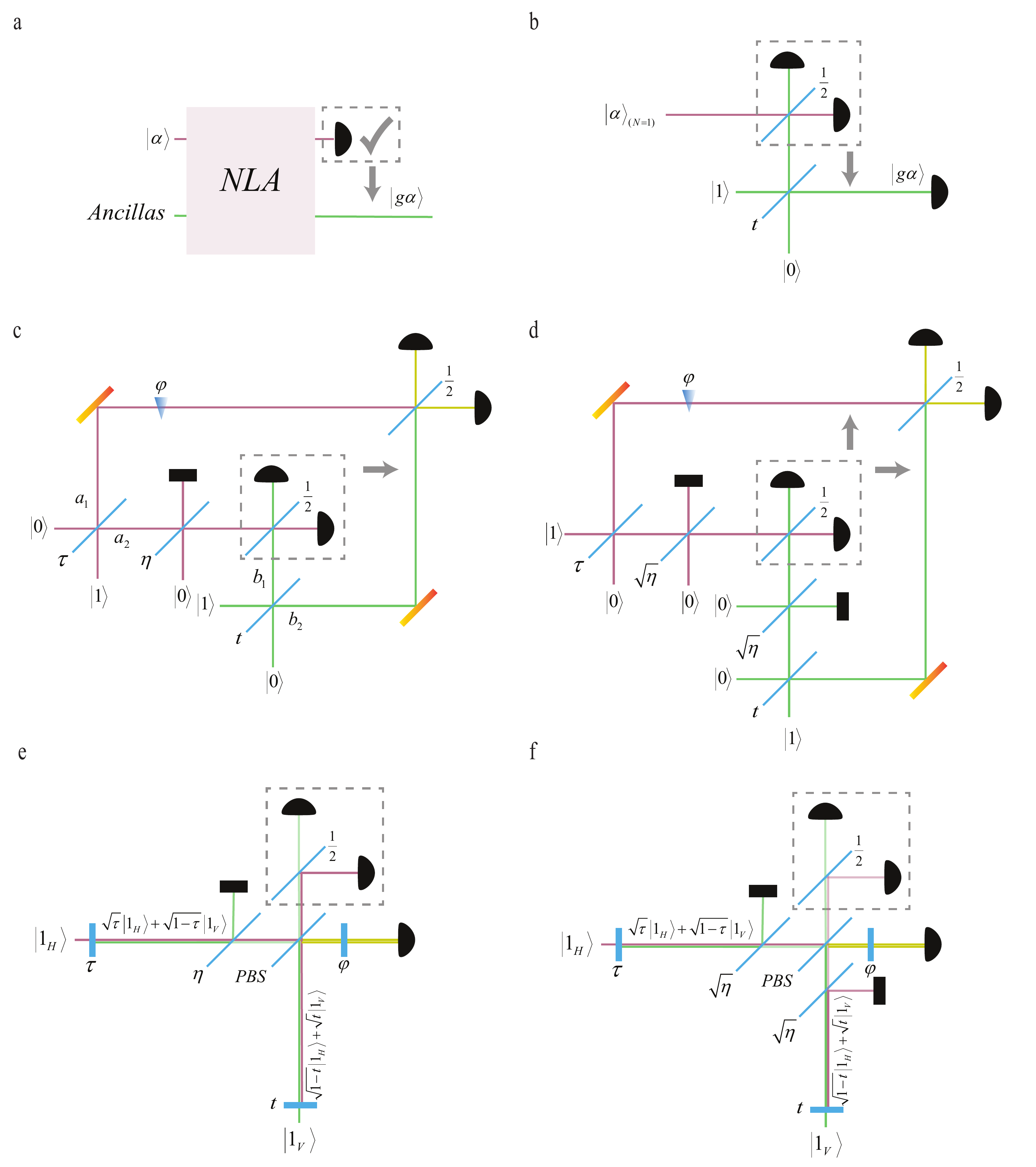}
	\caption{
(a) A schematic illustration of an NLA gate. An iput state $\ket{\alpha}$ is amplified to $\ket{g\alpha}$ if a correct signal is detected at the heralding detector of the NLA. An additional resource state is input to power the scheme. 
(b) A generalised quantum scissors scheme~\cite{Pegg1998}, that approximates the NLA for input states \( (1 - k)\left| 0 \right\rangle \left\langle 0 \right| + k \left| 1 \right\rangle \left\langle 1 \right| \). The state is amplified to \( (1 - k)\left| 0 \right\rangle \left\langle 0 \right| + k g^2 \left| 1 \right\rangle \left\langle 1 \right| \). Heralding the amplification is triggered by the detection of one photon in the heralding station (dashed box), provided \( g > 1 \).
(c) NLA characterisation in an entanglement distillation scenario: Alice, red color, prepares a single-rail entangled state in modes $ a_1$ and $a_2$, and sends one mode to Bob, with the shared state undergoing losses before reaching Bob, controlled by $\eta$. Bob prepares his ancilla entangled state in modes $ b_1$ and $b_2$, green color. Successful NLA event, heralded in the dashed box, flags sharing a distilled entangled state between Alice and Bob in modes $ a_1$ and $b_2$. Bob's output mode does not suffer any loss.
(d) Placing the heralding station halfway between Alice and Bob enhances the probability of successful NLA. Both Alice's and Bob's states suffer from loss controlled by $\sqrt{\eta}$.
(e, f) Dual-rail encoding of the schemes depicted in (c) and (d), using polarization states \( \left|1, 0 \right\rangle = \left|1_H\right\rangle \) and \( \left| 0,1 \right\rangle = \left|1_V\right\rangle \).
}
	\label{fig:concept}
\end{figure*}

An ideal NLA operation (see Fig.~\ref{fig:concept}a) transforms an input state in the Fock basis, 
\( \left| \alpha \right\rangle = \sum\limits_{n = 0}^N {{c_n}\left| n \right\rangle } \), 
into 
\( \left| g \alpha \right\rangle = \sum\limits_{n = 0}^N  {{g^n c_n}\left| n \right\rangle } \), 
where \( g \) is the amplitude gain of the NLA and a cut-off, $N$, is required for physical devices. For a heralded scheme, the success of the operation is typically heralded by a particular photon detection pattern at the heralding station (dashed box). 
For states with $N=1$, this transformation is achieved using the generalized quantum scissors scheme~\cite{Pegg1998}, shown in Fig.~\ref{fig:concept}b.
The gain, \( g = t/(1-t) \), is determined by the transmissivity of the beam splitter acting on the ancilla photon, \( \left| 1 \right\rangle \). 
The successful detection of a single photon in the heralding station (dashed box) signifies amplification of the input state, provided that \( g > 1 \). This paper focuses solely on this version of the NLA.

The NLA characterization is most effectively performed in an entanglement distillation scenario, as shown in Fig.~\ref{fig:concept}c. In this setup, Alice prepares a single-rail, mode-entangled state 
\( \left| \psi_{\rm in}  \right\rangle = \sqrt \tau  {\left| {10} \right\rangle _{{a_1}{a_2}}} + \sqrt {1 - \tau } {\left| {01} \right\rangle _{{a_1}{a_2}}} \)
and sends one of the modes to Bob. Before reaching Bob, the entangled state undergoes loss, modelled by a beam splitter with transmissivity \( \eta \). The lossy state is a mixed state with two components,  $|\psi_{0}\rangle  =\sqrt{(1-\eta)} \sqrt{(1-\tau)}  |00 \rangle_{{a_1}{a_2}}$ and $ |\psi_{1}\rangle  =\sqrt \tau  {\left| {10} \right\rangle _{{a_1}{a_2}}} + \sqrt \eta  \sqrt {1 - \tau } {\left| {01} \right\rangle _{{a_1}{a_2}}} $, reducing it to:
%
%
\begin{equation}
\label{NLA_output}
{\rho _{in}} = \frac{{{P_0}\left| {{\psi_0}} \right\rangle \left\langle {{\psi_0}} \right| + {P_1}\left| {{\psi_1}} \right\rangle \left\langle {{\psi_1}} \right|}}{{{P_0} + {P_1}}},
\end{equation}
 with $P_0=\langle\psi_0|\psi_0\rangle$ and $P_1=\langle\psi_1|\psi_1\rangle$.
To distil entanglement, Bob prepares an ancilla photon, which is transformed by the NLA’s beam splitter into 
\( \left| \psi  \right\rangle = \sqrt {1 - t} {\left| {10} \right\rangle _{{b_1}{b_2}}} + \sqrt t {\left| {01} \right\rangle _{{b_1}{b_2}}} \). If exactly one photon is detected at the heralding station and in the absence of loss in the circuit, the shared state between Alice and Bob is a mixed state with two components,  $|{{\psi^{'}}_{0}}\rangle  =
\sqrt{(1-t)} \sqrt{ (1-\eta)}  \sqrt{(1-\tau)}  |00 \rangle_{{a_1}{b_2}}$ and $|{{\psi^{'}}_{1}}\rangle  =\sqrt{1-t} \sqrt \tau  {\left| {10} \right\rangle _{{a_1}{b_2}}} +\sqrt\eta \sqrt {1 - \tau } \sqrt{t} {\left| {01} \right\rangle _{{a_1}{b_2}}} $, given by:
%
%

\begin{equation}
\label{NLA_output}
{\rho _{out}} = \frac{{{P_0}^{'}\left| {{\psi^{'} _0}} \right\rangle \left\langle {{\psi^{'} _0}} \right| + {P_1}^{'}\left| {{\psi^{'} _1}} \right\rangle \left\langle {{\psi^{'} _1}} \right|}}{{{P_0}^{'} + {P_1}^{'}}},
\end{equation}
\label{NLA_output}
with $P_0^{'}=\langle\psi_0^{'}|\psi_0^{'}\rangle$ and $P_1^{'}=\langle\psi_1^{'}|\psi_1^{'}\rangle$. The two modes carrying distilled entanglement, $a_1$ and $b_2$, can be then overlapped for quantum state characterization.

 We define \( p \) as the NLA success probability, i.~e. the probability of detecting a click in the heralding station. The fidelity of the output with the target state is given by $F=|\langle\psi_{\rm in}\left| \psi^{'}_{\rm 1} \right\rangle|^2$. At the same time, we use $X=\frac{P_1^{'}}{P_0^{'}}$, the ratio between the single-photon and vacuum components of the state, to represent the purity of the amplified state. The relationship between $X$ and the more standard definition of purity is given by $\textbf{Tr}({\rho ^2}) = \frac{{1 + {X^2}}}{{{{(1 + X)}^2}}}$. Ideally, an NLA configuration achieves a high success probability (\( p \)) while maximizing \( X \), thereby minimizing the vacuum component in the output state. While our primary focus is optimizing \( p \), whilst maintaining near maximal fidelity, we also evaluate and discuss the impact on \( X \) in this work.

The conventional application of the NLA occurs at Bob’s end as illustrated in Fig.~\ref{fig:concept}c. However, Mauron and Ralph~\cite{Mauron2022} suggest that positioning the heralding station mid-way between Alice and Bob, as shown in Fig.~\ref{fig:concept}d, significantly increases the probability of success~\cite{Mauron2022}. In this arrangement, the entangled states sent by Alice and Bob each pass through half of the channel, which has transmissivity \( \sqrt{\eta} \), before interfering at the heralding station. While the overall transmissivity remains \( \eta \), this midpoint configuration enhances the scaling of success probability by a factor of approximately \( \frac{1}{2\sqrt{\eta}} \), for large $X$. In Fig.~\ref{fig:concept}e and f, we show the polarisation encoded version of the schemes from Fig.~\ref{fig:concept}c and d, implemented with polarization states \( \left| 1,0 \right\rangle = \left|1_H\right\rangle \) and \( \left| 0,1 \right\rangle = \left|1_V\right\rangle \). This implementation enables the spatial overlap of output modes \( a_1 \) and \( b_2 \), facilitating analysis of the final state using polarization-based quantum state tomography techniques.

\section{Experiment}

\begin{figure}
	\centering
	\includegraphics[width=0.90\linewidth]{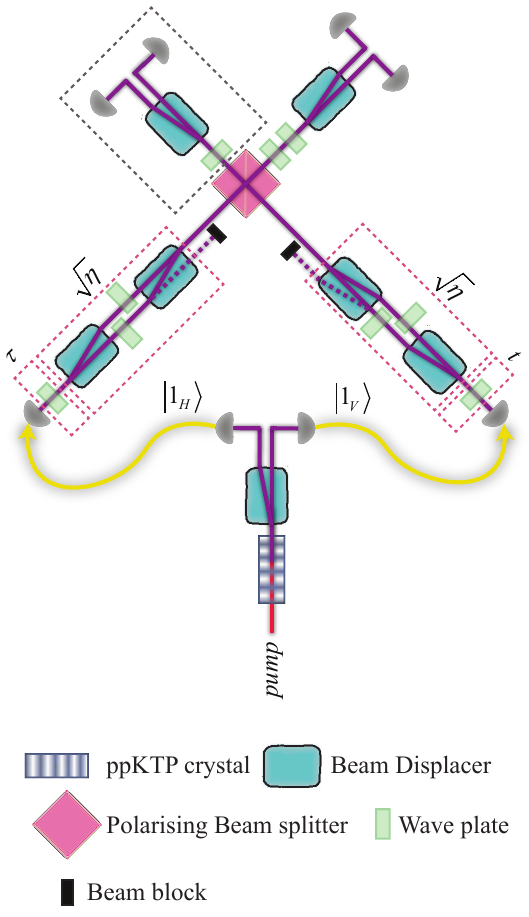}
	\caption{
 The experimental setup: Photon pairs are generated through an engineered spontaneous parametric down-conversion (SPDC) process. These photons are sent to the optical setup, which implements the schemes shown in Figs.~\ref{fig:concept}e-f. The parameters \(t\) and \(\tau\) are controlled using two wave plates. Transmissivity ($\sqrt{\eta}$ is adjusted in two arms via Mach-Zehnder interferometers, each implemented by two alpha-BBO polarizing beam displacers. D-shaped waveplates allow fine-tuning of the transmissivity in one polarisation basis, from 0 to 1, without affecting the orthogonal polarisation. The input and reference beams interfere at the central polarising beam splitter (PBS), and the photons are ultimately detected by superconducting nanowire single-photon detectors and counting modules.
}

	\label{fig:experiment}
\end{figure}

Fig.~\ref{fig:experiment} illustrates the experimental setup. Heralded single photons are generated via engineered spontaneous parametric down-conversion (SPDC). A 15 mm periodically poled potassium titanyl phosphate (PPKTP) crystal is pumped by a pulsed $775~\rm{nm}$ Ti-sapphire laser, producing spectrally degenerate photons at $1550~\rm{nm}$. The photon source is designed~\cite{weston2016efficient} to generate spectrally pure and indistinguishable photons with high heralding efficiency~\cite{klyshko1980use}, both being critical requirements for the correct operation of the NLA.

Heralding efficiency strongly influences the purity of the amplified state and the probability of success of the NLA. Lower efficiency on the $b_2$ mode (see Fig.~\ref{fig:concept}c) increases the likelihood of detecting a heralding event without the corresponding output photon, which reduces the final state's purity. Spectra-temporal indistinguishability between the signal and idler photons of SPDC is essential for high-quality quantum interference at the heralding station of the quantum scissors gate. Imperfect interference increases the error rate of the NLA, contributing to noise in both the heralding station and the output mode of the NLA. Our source achieved a heralding efficiency above $80\%$, including detection efficiency. At the same time, we observed above $99\%$ interference visibility within the NLA without employing any spectral filtering of the photons. 

Another critical factor is the dark counts (the average rate of registered photon counts without any incident light) of the detectors in the heralding station. These can cause false heralding events, thereby reducing the purity of the output state. This issue becomes more pronounced in the low-transmissivity and high-gain regime, where the number of real and dark counts is comparable. In this regime, spurious heralding events significantly degrade the output purity. To mitigate this effect, we gate the detectors with a 700 ps coincidence window, triggered by the pump laser pulse. This reduces the dark counts to around 100 per second. The photons are ultimately detected by superconducting nanowire single-photon detectors and counting modules.

The generated photons are directed into the optical setup, as shown in Fig.~\ref{fig:experiment}, which implements the schemes depicted in Fig.~\ref{fig:concept}e and f. The polarisation mode-dependent loss is applied via a Mach-Zehnder interferometer, implemented with polarizing beam displacers. One of the polarisation modes propagates without loss through the interferometer, while the orthogonal mode is partially redirected to a different output using a half-wave plate before the second beam displacer. The use of low-loss beam displacers and D-shaped waveplates allows for the unaffected mode to propagate without incurring unwanted loss while maintaining high interferometric stability of the setup and the ability to fine-tune the parameters \( t \) and \( \tau \) by adjusting the corresponding wave plates. The modes affected by loss are interfered with at the central beam splitter of NLA and sent toward the heralding station, while the modes unaffected by loss are propagated directly to the output detectors and analysed via photon counting and polarization quantum state tomography.

\begin{figure*}[tb]
	\centering
\includegraphics[width=1\linewidth]{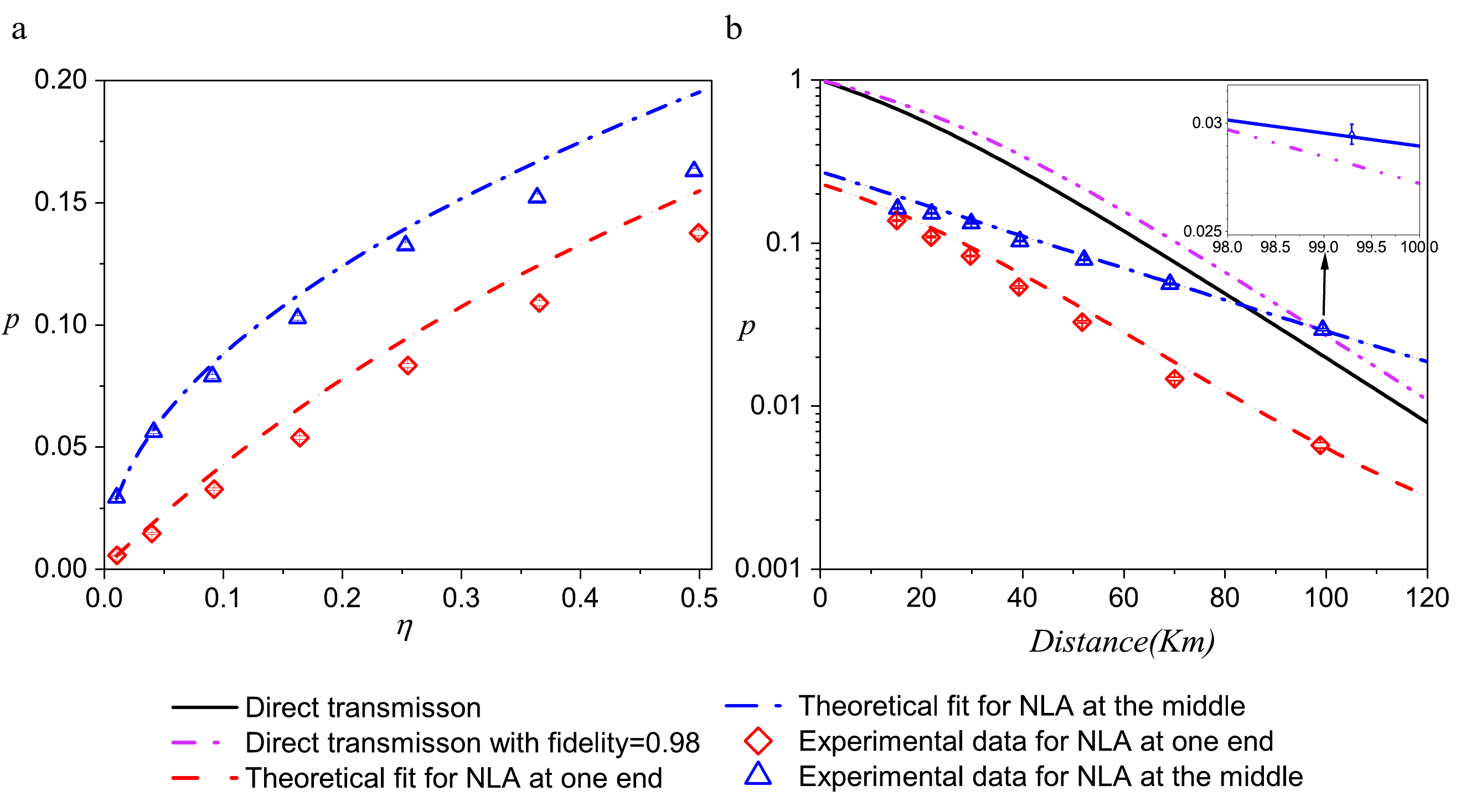}
	\caption{The probability of success, \(p\), (a) as a function of transmissivity, \(\eta\), and (b) distance in fibre, for the case that the input state is fixed to the maximally entangled state \(D=\frac{1}{\sqrt{2}} \left( \left| 1_H \right\rangle + \left| 1_V \right\rangle \right)\).  The results indicate that placing NLA in the middle increases the probability of success. When it is compared to direct transmission in fibre with 0.2 dB/Km loss, NLA in the middle outperforms a direct transmission channel with fidelity of 0.98. The theoretical lines take into account source and detector losses, and dark counts. Experimental uncertainties are estimated based on Poissonian photon statistics, with corresponding error bars smaller than the plot markers. (See Methods for the details on the model and exact values of fit parameters.)}
	\label{fig:result1}
\end{figure*}

\section{Results}

We first analyze the performance of the NLA in a distillation scenario when a maximally entangled input state corresponding to $\tau=0.5$, affected by loss, is distilled into a maximally entangled state. For this purpose, we prepare our input state $\left|\psi_{\rm in}\right\rangle$ as $\left|{D} \right\rangle=\frac{1}{{\sqrt 2 }}(\left| {{1_H}} \right\rangle  + \left| {{1_V}} \right\rangle )=\frac{1}{{\sqrt 2 }}(\left| {1,0} \right\rangle  + \left| {{0,1}} \right\rangle )$. We then record the probability of success $p$ of the NLA for the gain setting $t$, which is theoretically supposed to yield a maximally entangled state in the output. The corresponding setting satisfies $t=\frac{\tau}{\eta + \tau - \tau \eta}$ and $t=\tau$ for NLA at one end and NLA at the middle, respectively. The fidelity of the output state is quantified by $F(\rho, \sigma) = \left[ \mathrm{Tr}\sqrt{\sqrt{\rho}\,\sigma\sqrt{\rho}} \right]^2
$, where $\rho=|\psi^{'}_1\rangle\langle \psi^{'}_1|$ and $\sigma=|D\rangle \langle D|$. 

Figure~\ref{fig:result1}a shows the probability of success, $p$, for the two cases of NLA positioned either at the end or in the middle of the lossy channel of total transmission $\eta$ ranging between $0.01$ and $0.49$. We achieve high purity and fidelity of the amplified state, with an average fidelity $F=0.956 \pm 0.038$ and an average purity $X_{\rm min}=0.557\pm 0.098$. As it can be seen, positioning the NLA in the middle of the channel provides a significant advantage in terms of $p$, over the case when the NLA is positioned at the end of the channel. We have theoretically modeled the experiment incorporating realistic effects including source and detector inefficiencies and dark counts. See Methods for more details on the theoretical model.

To put our results in more practical terms, we compare the probability of success vs communication distance in a standard fibre with $0.2$ dB/Km loss. As demonstrated in Fig.~\ref{fig:result1}b, the two NLA schemes show different scaling with distance, where NLA at the middle scales with $\sqrt{\eta}$ and clearly outperforms the other strategy. More significantly, we show that in the high loss regime ($\sim$ 100 Km), the NLA in the middle also overcomes the probability of success of a direct transmission channel with the same loss. The direct transmission line in Fig.~\ref{fig:result1}b is the `do nothing' protocol from Ref.~\cite{Mauron2022}. To make a fair comparison, we have considered a case where direct transmission has fidelity (0.98), which is comparable to the fidelity that our experiment achieves in this regime. 

\begin{figure}[tb]
	\centering
\includegraphics[width=1\linewidth]{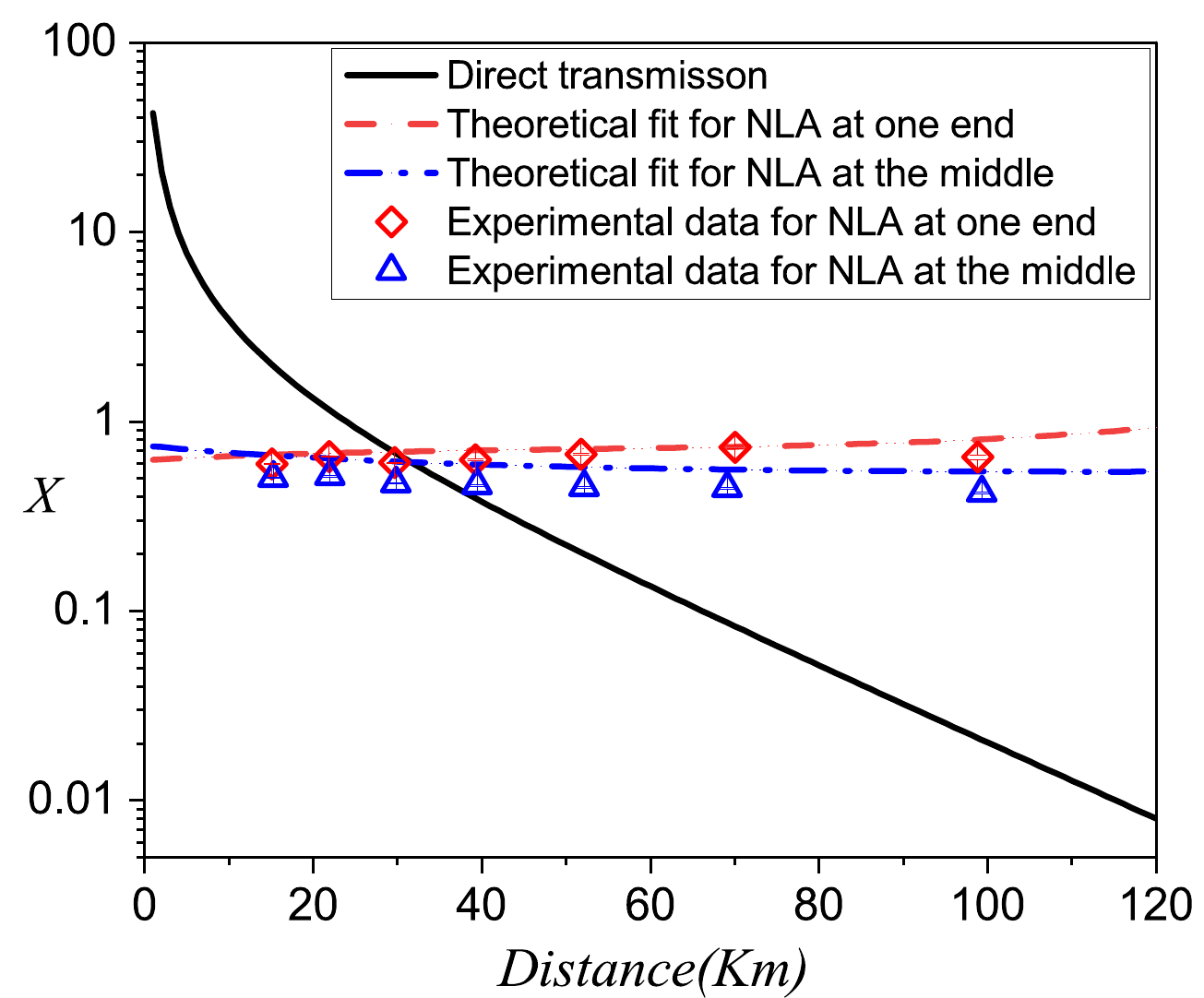}
	\caption{The purity of the output state $X$ as it scales with distance. The solid black line indicates the purity without the NLA. Both NLA schemes are able to preserve purity in the presence of high photon loss. The theoretical lines take into account source and detector losses, and dark counts. Experimental uncertainties are estimated based on Poissonian photon statistics.}
	\label{fig:result2}
\end{figure}

In Fig.~\ref{fig:result2}, we compare the purity of the distilled state, $X$ vs distance. One can see that, in direct transmission, $X$ drastically drops with distance, while in NLA the purity hovers around a fixed value. Substantially, in the high loss regime, the relative increase in $p$ does not decrease the purity of the state. The strength of this scheme is to preserve the purity $X$ in the presence of high photon loss, as demonstrated here .

Further improvement of the NLA operation can be achieved by tuning the gain. To account for experimental imperfections, we allow the gain parameter (controlled by $t$), to slightly deviate from the nominal theoretical values. That is, the input state is set to the maximally entangled state, and the gain setting is tuned to provide an output state $\left|\psi_{\rm out} \right\rangle$ closest to $\left|{D} \right\rangle$, quantified by fidelity. The results corresponding to this scenario are shown in supplementary material (See Supplementary Figure 2), demonstrating a more pronounced improvement in $p$ for the case of NLA positioned in the middle of the channel. Furthermore, in this case, we achieved even higher fidelity (mean $F=0.968\pm 0.020$) and purity (mean $X=0.602\pm0.074$ ) of the amplified state.

While in both cases, positioning NLA in the middle of the channel proves to be advantageous in terms of the probability of success, the experimental results slightly deviate from the expected behaviour, specifically for the very high transmissivity of the channel. The deviation in the high transmissivity regime is expected because the quantum scissor-based NLA does not achieve its ideal operation in this regime~\cite{ralph2008nondeterministic}. 
The observed deviation from the anticipated scaling of \( p \) in the low \( \eta \) regime is attributed to the influence of dark counts at the heralding station, which becomes comparable in magnitude to true heralding events within this range.


\section{Discussion}

In summary, we experimentally demonstrated a quantum state amplification approach with an increased probability of success. The advantage is achieved without detrimental impact on the purity and fidelity of the amplified state. We show different scaling of the probability of success, for two different schemes, NLA at one end and NLA at the middle. Our experiment further highlights the stringent experimental requirements for the operation of NLA. Besides the necessarily high efficiency of photon delivery, we observe a strong influence of dark counts at the heralding stations on the purity of the output state. Our results comprehensively report different numbers crucial for NLA, keeping high fidelity, achieving a high probability of success, and high purity. Nevertheless, our experimental demonstration matches well the theoretical prediction of improved scaling of the probability of success for a large range of channel loss. The ability to achieve a higher probability of success of an amplification or distillation protocol will affect the design choices of the quantum repeater architecture and sets new conditions on the performance of corresponding components, such as quantum memories and quantum state sources. Our work provides a strong insight into the future design of long-distance quantum communication schemes and quantum repeaters, and highlights the importance of experimental imperfections.

Considering that NLA schemes improve the probability of success while preserving the purity, one significant experimental challenge for future works is to improve the purity of the transferred state. This is directly related to the efficiency of the source, the setup and the detectors. In our case, in the ideal scenario, $X$ is upper bounded to $\frac{\textbf{heralding efficiency}}{1- \textbf{heralding efficiency}}$. Therefore, a significant improvement can be achieved in  $X$, by a modest improvement of the heralding efficiency from the state-of-the-art number in current technology ($\approx 60 \% $).

\section{Authors Contributions}
S.S., G.J.P., and T.C.R. conceived the experiment. F.G. and S.S. designed the experimental setup. F.G. performed the experiment with assistance from S.S., L.K.S. and V.B.V. J.D. and T.C.R. did the theory work. F.G., S.S., T.R., and J.D. analyzed the experimental results. All authors contributed to writing the manuscript.
\section{Acknowledgments}

F.G was supported partly by the Griffith University Postdoctoral Fellowship (GUPF\#58938).

\section{Methods}
\subsection{Accounting for experimental imperfections}
To capture the effects of realistic components in our experiment, we have modelled our protocol incorporating imperfect sources and detectors. Theoretical work by Mauron and Ralph models inefficiency in some elements of this experiment \cite{Mauron2022}. Here, we expand on their analysis for a comprehensive theoretical model incorporating inefficiency in all sources and detectors, as well as detectors with dark counts and without photon-number resolution. In Supplementary Figure 1, we illustrate how each realistic element was modelled in the NLA halfway protocol. Alice and Bob’s source inefficiency is modelled via a beam splitter of transmissivity $\epsilon_1$ and $\epsilon_2$ immediately after their respective single-photon sources. The four detectors used in the protocol are characterized by inefficiency $\delta_1$ for the two NLA heralding detectors and $\delta_2$ for the two state characterization detectors. These inefficiencies are modelled via a beam splitter before each detector. Lastly, detector dark counts are modeled via clicks at fictitious detectors.

Through the above analysis, we have fitted a model to our experimental data, yielding the following detection efficiencies: ${\delta}_1 \approx 0.95, {\delta}_2 \approx 0.80$, with source efficiency $\epsilon_1 =\epsilon_2 \approx 0.85$ for the NLA in the middle protocol and $\epsilon_1 =\epsilon_2 \approx 0.78$  for the NLA at one end scheme in Fig~\ref{fig:result1}. These parameters correspond to a heralding efficiency ranging from $68\%-80\%$ for the NLA in the middle protocol and $62\%-74\%$ for the NLA at one end scheme. These parameters are consistent with our experimental values and demonstrate the quality of the experimental implementation.

\section{Supplementary Information: Entanglement distillation rates exceeding the direct transmission bound}



\begin{figure*}
	\centering
\includegraphics[width=0.5\linewidth]{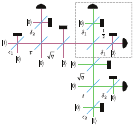}
	\caption{Theoretical model of the NLA at the middle protocol in the presence of imperfect source and detection components. The efficiency of the NLA heralding detectors and state characterization detectors is parametrised by $\delta_1$ and $\delta_2$, respectively. Alice’s source efficiency is $\epsilon_1$ and Bob’s source efficiency is $\epsilon_2$. Detector dark counts are modelled via a thermal state input into the environmental mode of the beam splitter modelling detection inefficiency.}
	\label{SIfig:detection-ineff}
\end{figure*}

\begin{figure*}[tb]
	\centering
\includegraphics[width=1\linewidth]{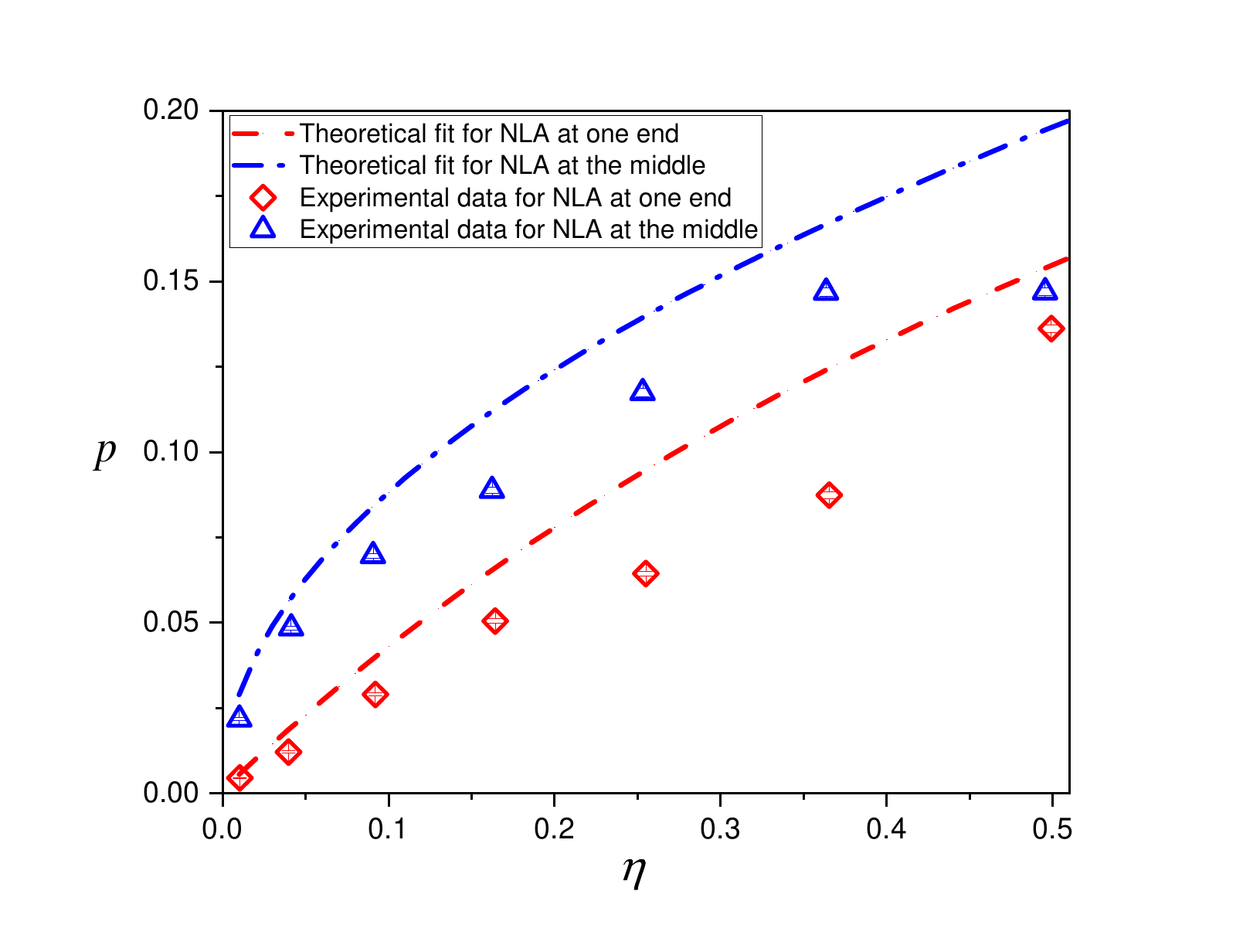}
	\caption{  The probability of success, \(p\), as a function of transmissivity, \(\eta\).The performance of the NLA operation can be further enhanced by fine-tuning the gain. To address experimental imperfections, the gain parameter (controlled by $t$) is allowed to deviate slightly from its ideal theoretical values. In this approach, the input state is prepared as a maximally entangled state and the gain is adjusted to produce an output state $\left|\psi_{\rm out} \right\rangle$ with maximum fidelity to the target state $\left|{D} \right\rangle$. The advantage of positioning the NLA in the middle of the lossy channel is evident, with the difference is more prominent compared to Fig. 3a, where we observe an increase in the output state fidelity, see main text for details. The theoretical lines take into account source and detector losses, and dark counts. Experimental uncertainties are estimated based on Poissonian photon statistics.}
	\label{SIfig:max_F}
\end{figure*}
\end{document}